\documentclass[letterpaper,english,aps,prl,twocolumn,floatfix,showpacs,amsfonts,amssymb]{revtex4}

\usepackage[T1]{fontenc}
\usepackage[latin1]{inputenc}
\usepackage{graphics}
\usepackage{amssymb}
\usepackage{amsmath}
\usepackage{graphicx,epsf}

\newcommand{\beq}{\begin{equation}}
\newcommand{\eeq}{\end{equation}}
\newcommand{\bea}{\begin{eqnarray}}
\newcommand{\eea}{\end{eqnarray}}
\newcommand{\eps}{\epsilon}
\begin{document}
\def\tende#1{\,\vtop{\ialign{##\crcr\rightarrowfill\crcr
\noalign{\kern-1pt\nointerlineskip} \hskip3.pt${\scriptstyle
#1}$\hskip3.pt\crcr}}\,}

\title{Restoration of the magnetic hc/e -periodicity in
unconventional superconductors}

\author{Vladimir  Juri\v ci\' c$^1$, Igor F. Herbut$^1$,
and Zlatko Te\v sanovi\' c$^2$}

\affiliation{$^1$ Department of Physics, Simon Fraser University,
 Burnaby, British Columbia, Canada V5A 1S6 \\
 $^2$ Department of Physics and Astronomy, Johns Hopkins University, Baltimore, MD 21218, USA}

\begin{abstract}
We consider the energy of the filled quasiparticle's Fermi sea of
a macroscopic superconducting ring threaded by an
$hc/2e$-vortex, when the material of the ring is of an
unconventional pairing symmetry. The energy relative to
the one for the $hc/e$-vortex configuration is finite, positive,
and inversely proportional to ring's inner radius. We argue that
the existence of this energy in unconventional superconductors
removes the commonly assumed degeneracy between the odd and the
even vortices, with the loss of the concomitant $hc/2e$
periodicity in external magnetic field as a consequence.
This macroscopic quantum effect should be observable in nanosized
unconventional superconductors with a small phase
stiffness, such as deeply underdoped YBCO with $T_c < 5K$.

\end{abstract}

 \maketitle

A fundamental property of all known superconductors is that
their electrons form Cooper pairs. A direct
manifestation of this phenomenon is the quantization of magnetic flux
in units of $hc/2e$  in multiply connected
geometries \cite{doll}. Such flux quantization may be considered as an effective
spectroscopy of the charge of the carriers, and is often used
as a proof of paired nature of the superconducting state.
A closely related phenomenon is the periodicity of various properties of multiply
connected superconductors in the external magnetic field with the period
that corresponds to the half flux quantum \cite{little}.

In this Letter we argue that this periodicity is in principle {\em
not exact} in superconductors with unconventional pairing symmetry
that support quasiparticle excitations at {\em arbitrarily low}
energies. Our argument is qualitative and fundamental in nature,
and based on a calculation of the difference in ground state
energies of nodal quasiparticles of an unconventional
superconductor in a presence of a single ($hc/2e$) and a double
($hc/e$) vortex in an annular geometry. An estimate of this energy
difference indicates that the deviations from the familiar $hc/2e$
periodicity may become directly observable in recently fabricated
deeply underdoped cuprates. This manifestation of macroscopic
quantum coherence is a fundamental effect and raises the
possibility of manipulation of spin currents (carried by
quasiparticles) by controlled motion of magnetic fluxes (using,
for example, a Hall bar) with applications to spintronics and
other areas of applied science.

The energy in question stems from the essential difference in the
interactions between the quasiparticles and $hc/e$- and
$hc/2e$-vortices. Apart from the semiclassical Doppler shift of
the quasiparticle energies common to both single and double
vortices, the statistical, purely {\em quantum} Aharonov-Bohm (AB)
phase of an $hc/e$-vortex can be exactly gauged away whereas the
one of an $hc/2e$-vortex cannot. The ensuing topological
frustration is felt by the quasiparticles arbitrary far from the
center of the vortex via the AB gauge field \cite{FT}, which
encodes the sign change in the quasiparticle's wave-function as it
is adiabatically dragged around the vortex. Vortices, their
fluctuations and the concomitant AB  and Doppler effects on
quasiparticles in d-wave superconductors have been a subject of
much research in the past
\cite{Melnikov,marinelli,ashvin,franz,herbut,dlee,Nikolic,MT,index}.
Here we consider the filled Fermi sea of nodal quasiparticles in
an annular geometry (Fig. 1), and determine the excess in energy
due to the AB gauge field of the $hc/2e$-vortex. We find a
positive contribution to the condensation energy that derives
predominantly from the quasiparticles near the nodes and is
inversely proportional to the hole radius R. For parameters
relevant to cuprates the excess energy is $\sim 0.2K$ for a thick
ring whose inner radius is about a {\em micrometer}. Consequences
for the quantization of magnetic flux in underdoped cuprates are
briefly discussed.

 \begin{figure}
\begin{center}
\includegraphics[width=3cm]{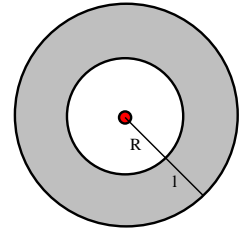}
\end{center}
\caption{ (Color online.) Annular system with the vortex in the nodal superconductor (gray region). }
\label{Fig-annulus}
\end{figure}

Let us assume a magnetic flux localized in the annulus made
of an unconventional
superconductor. At low-energies, the dynamics of quasiparticle excitations near a single node
in the field of a vortex in the superconducting order parameter carrying a half flux quantum
$hc/2e$ may be described by the Hamiltonian
\begin{equation}\label{Hamiltonian}
{\hat H}=v_F(p_x+a_x)\sigma_1+v_\Delta(p_y+a_y)\sigma_2-m\sigma_3,
\end{equation}
where $v_F$ and $v_\Delta$ are characteristic velocities of the
quasiparticle excitations in the two directions around a nodal
point, $\sigma_i$ are
the Pauli matrices. ${\bf a}({\bf r})=(x{\bf e}_y-y{\bf e}_x)/(2r^2)$
is the AB vector potential resulting from the Franz-Te\v sanovi\' c \cite{FT} (FT)
gauge transformation in presence of the $hc/2e$-vortex in the order parameter.
In (\ref{Hamiltonian}) we have included a gap
$m$ at the nodes for generality and set $\hbar=c=1$. We have also
neglected the Doppler shift of quasiparticles, based on the following
argument: the Doppler effect enters Hamiltonian (\ref{Hamiltonian}) as
another gauge field ${\bf v}({\bf r})$ generated by the FT
transformation. It always appears in the
combination ${\bf v}({\bf r})- (e/c){\bf A}({\bf r})$, where
${\bf A}({\bf r})$ is the electromagnetic vector potential. When the hole
radius $R$ (Fig. 1) grows to become comparable to the magnetic
field penetration depth $\lambda$, ${\bf A}({\bf r})$ will {\em screen} out
${\bf v}({\bf r})$, leaving the topological
frustration encoded by ${\bf a}({\bf r})$ as the sole
long range effect. Thus, in practical situations, we expect the Doppler
shift to be a secondary effect in rings of macroscopic size.

We consider first
the simpler case of isotropic velocities,
$v_F=v_\Delta$. Setting the velocity to unity the eigenstates of
the above Hamiltonian are found to be
\begin{eqnarray}\label{eigenstates}
\Psi_{q,k,l}(r,\phi)&=&\sqrt{\frac{k}{4\pi
|E|}}\nonumber\\
&\times&\left(\begin{array}{cc}
\sqrt{q(E-m)}J_{l-1/2}(k r)e^{i(l-1)\phi}\\
iq\sqrt{q(E+m)}J_{l+1/2}(k r)e^{i l\phi}
\end{array}\right)
\end{eqnarray}
for $l>0$, and
\begin{eqnarray}\label{eigenstates1}
\Psi_{q,k,l}(r,\phi)&=&\sqrt{\frac{k}{4\pi
|E|}}\nonumber\\
&\times&\left(\begin{array}{cc}
\sqrt{q(E-m)}J_{-l+1/2}(k r)e^{i(l-1)\phi}\\
-iq\sqrt{q(E+m)}J_{-l-1/2}(k r)e^{i l\phi}
\end{array}\right)
\end{eqnarray}
for $l\leq 0$. Here the quantum number $q=\pm1$ distinguishes
particle-like and hole-like states, $l\in{Z}$ is angular momentum,
$k>0$ is radial wave vector, and energy is $E_{q,
k}=q\sqrt{k^2+m^2}$. $J_l(x)$ are Bessel functions of the first
kind. To find the above eigenstates it is necessary to regularize
the gauge potential. Namely, the requirement of the
square-integrability unambiguously determines all the eigenstates
of the Hamiltonian (\ref{Hamiltonian}), except ones with zero
angular momentum. There are two $l=0$ states diverging as
$1/\sqrt{r}$ at the origin, but keeping both of them leads to an
overcomplete basis in the Hilbert space. On the other hand,
requirement of non-divergence of the states at the origin is too
restrictive, since it leads to an incomplete eigenbasis. Only
linear combinations of the two states specified by a single
parameter are allowed \cite{GJ} but in order to select a single
eigenstate from all allowed states, the gauge potential has to be
regularized. Here, we considered the vortex as a cylinder of a
finite radius $R$ in which the AB flux is uniformly distributed.
By matching the solutions inside and outside the cylinder, and
taking the limit $R\rightarrow0$, we found that the eigenstate in
the zero angular momentum channel has the form given by
(\ref{eigenstates1}), with the lower component diverging at the
origin. This is in agreement with the result of an alternative
regularization \cite{MT,GJ}.

Let us now calculate the local density of states (LDOS) for
gapless nodal quasiparticles, $m=0$, defined as
\begin{equation}
\rho(\epsilon,{\bf r})=\sum_{q,l}\int
dk|\Psi_{q,l,k}|^2\delta(\epsilon-E_{q,k}).
\end{equation}
Using the eigenstates given by Eqs.\ (\ref{eigenstates}) and
(\ref{eigenstates1}), we reproduce the LDOS of Ref.\ \onlinecite{Nikolic} in the form
\begin{equation}
\rho(\epsilon,{\bf
r})=\frac{\cos(2|\epsilon|r)}{2\pi^2r}+\frac{|\epsilon|}{\pi}\sum_{l=0}^\infty
J_{l+1/2}^2(|\epsilon|r).
\end{equation}
The expression for the LDOS can further be simplified using the little known
Mitrinovi\' c identity \cite{Hansen}
\begin{equation}
\sum_{l=1}^\infty [J_{p+l}(x)]^2=p\int_0^x\frac{dt}{t}
[J_p(t)]^2-\frac{1}{2}[J_p(x)]^2.
\end{equation}
For $p=1/2$, this yields the form of the LDOS which is more convenient for
the later calculations
\begin{equation}\label{LDOS}
\rho(\epsilon,{\bf
r})=\frac{1}{\pi^2}\left[\frac{\cos(2|\epsilon|r)}{2r}+|\epsilon|{\rm
Si}(2|\epsilon|r)\right],
\end{equation}
where the standard sine-integral function is defined as ${\rm
Si}(x)\equiv\int_0^x dt \sin t/t$. In the vicinity of the vortex,
in the region $r\ll 1/|\epsilon|$, the LDOS diverges as $1/r$.
This behavior of the LDOS originates from the states in the zero
angular momentum channel that diverge as $1/\sqrt{r}$, when
$r\rightarrow0$. On the other hand, far from the vortex,
$\rho(\epsilon,{\bf r})\rightarrow\rho_0(\epsilon,{\bf
r})=|\epsilon|/2\pi$, as in the vortex-free system. Of course, the LDOS in the
system with the vortex carrying an integer number of the flux
quanta, $n hc/e $, $n\in Z$, is the same as in the free
system. Namely, the vector potential corresponding to $n hc/e $-vortex
is $2n{\bf a}({\bf r})$, and in that case the eigenstates of
(\ref{Hamiltonian}) have the form
\begin{eqnarray}
\Psi_{q,k,l}(r,\phi)&=&\sqrt{\frac{k}{4\pi
|E|}}\nonumber\\
&\times&\left(\begin{array}{cc}
\sqrt{q(E-m)}J_{|l-1+n|}(k r)e^{i(l-1)\phi}\\
iq\sqrt{q(E+m)}J_{|l+n|}(k r)e^{i l\phi}
\end{array}\right).
\end{eqnarray}
The LDOS for gapless nodal quasiparticles $\rho_0 (\epsilon, {\bf
r})$ is then uniform and independent of the integer $n$, as
required by gauge invariance.

 Starting with the compact form of the LDOS, we may
compute the energy cost of having an $hc/2e$-vortex
 by integrating the Eq.\ (\ref{LDOS}) over the energy and the area of the ring in Fig. 1.
 This procedure should be accurate for a ring of a macroscopic size,
 when the effects of the boundaries and of the discreetness of the spectrum become negligible.
 The DOS for the ring is then
\begin{equation}
\rho(\epsilon)=\int d^2{\bf r}\rho(\epsilon,{\bf
r})=I(R+l,\epsilon)-I(R,\epsilon),
\end{equation}
where
\begin{equation}
I(R,\eps)\equiv\frac{\eps R^2}{\pi}\left[{\rm Si}(2\eps
R)+\frac{\cos(2\eps R)}{2\eps R}+\frac{\sin(2\eps R)}{(2\eps
R)^2}\right].
\end{equation}
$R+l$ and $R$ are the radii of the outer and the inner annulus,
respectively. The total energy of the system then becomes
\begin{equation}\label{totalenergy1}
{\cal E}=-\int_0^\Lambda d\eps\,\, \eps \rho(\eps)={\tilde{\cal
E}}(R+l)-{\tilde{\cal E}}(R),
\end{equation}
with
\begin{eqnarray}
{\tilde{\cal E}}(R)&=&-\frac{\Lambda^3R^2}{3\pi}\left[{\rm
Si}(2R\Lambda)+\frac{\cos(2R\Lambda)}{2R\Lambda}+\frac{\sin(2R\Lambda)}
{(2R\Lambda)^2}\right.\nonumber\\
&+&\left.\frac{1-\cos(2R\Lambda)}{4R^3\Lambda^3}\right].
\end{eqnarray}
Here, $\Lambda$ is a high-energy cutoff, and the minus sign in
Eq.\ (\ref{totalenergy1}) takes into account that only hole-like
states are occupied in the ground state. Using the asymptotic form
of the sine-integral function for large values of its argument, we
find the energy cost of an $hc/2e$-vortex in the annulus of a
macroscopic size $R\gg1/\Lambda$ to be
\begin{equation}\label{vortexE-isotropic}
{\cal E}_v\equiv{\cal E}-{\cal E}_0=\frac{l}{12\pi R(R+l)} \left( 1+{\cal
O}(\frac{1}{\Lambda R}) \right),
\end{equation}
where the total energy for the $hc/e$-vortex (or the vortex-free
system) is ${\cal E}_0=\Lambda^3[R^2 - (R+l)^2]/6$. When the
thickness of the annulus is much larger than its inner radius,
$l\gg R$, the extra energy cost due to the presence of an
$hc/2e$-vortex in the order parameter, to the leading order in
$1/\Lambda R$ and $R/l$, is simply
\begin{equation} \label{simple}
{\cal E}_v=\frac{\hbar v_F}{12\pi R},
\end{equation}
where we have also restored Planck's constant and the Fermi
velocity $v_F = v_\Delta$ previously set to one. The energy cost
for having an $hc/2e$-vortex in the system therefore is {\it
positive}, and in the macroscopic limit, for a thick annulus,
inversely proportional to its inner radius. Notice that the
leading term in ${\cal E}_v$ is independent of the high-energy
cutoff $\Lambda$, in accord with our assumption that the effect is
due to the low-energy quasiparticles near the nodes. The
long-wavelength, linearized, description we postulated in Eq. (1)
is thus internally consistent for an annulus of a macroscopic
size. The result in Eqs.\ (\ref{vortexE-isotropic}) and \
(\ref{simple}) also reflects the fact that the presence of the
vortex affects the LDOS in its (macroscopic) vicinity the most.

\begin{figure}
\begin{center}
\includegraphics[width=8cm]{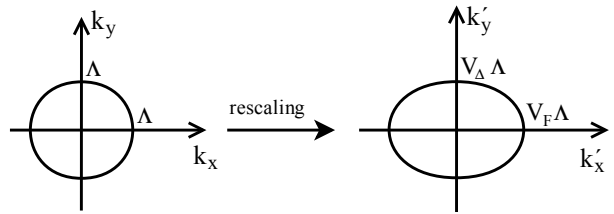}
\end{center}
\caption{Shape of the Brillouin zone before and after
rescaling of the momenta.} \label{Fig-brillouin}
\end{figure}

We can now turn to the general and a physically more relevant case when the two
characteristic velocities of the nodal quasiparticles, $v_F$ and
$v_\Delta$, are different. By rescaling the coordinates,
$x'=x/v_F$, $y'=y/v_\Delta$, and choosing a gauge such that the vector potential has the
same form as below Eq. (1) in the new coordinates $(x',y')$,  Hamiltonian
(\ref{Hamiltonian}) may be transformed to a form with isotropic
velocities \cite{Melnikov,Nikolic,MT}. The rescaled momenta are now
$k'_x=v_F k_x$, $k'_y=v_\Delta k_y$, and the dispersion assumes an
isotropic form, $E_{q,k'}=qk'$, with
$k'\equiv\sqrt{k_x^{'2}+k_y^{'2}}$. The LDOS then becomes $\rho(\epsilon, {\bf r'})/2\pi$,
with the extra factor of $2\pi$ arising from the elliptic shape
of the Brillouin zone in the new coordinates, shown in Fig.\
\ref{Fig-brillouin}.

The total energy of the anisotropic vortex-free system in the
annular geometry is then ${\cal E}_{0}=G(R+l)-G(R)$,  where
\begin{equation}\label{G}
G(R)\equiv-\int_{r\leq R} d^2{\bf r}\int_{\Sigma(\Lambda)}d^2{\bf
k'}\rho_0(k').
\end{equation}
$\Sigma(\Lambda)$ is the Brillouin zone in Fig. 2, and
$\rho_0(k')=k'/(2\pi)^2$ is the LDOS of the vortex-free system.
The integration yields the total energy of the
flux-free system
\begin{eqnarray}\label{vortex-free-anisotropic}
{\cal
E}_0&=&-\frac{\Lambda^3}{12}\frac{v_\Delta^2}{v_F}\left[(R+l)^2-R^2\right]
\left[_2F_1\left(\frac{1}{2},\frac{3}{2};1;1-\frac{v_\Delta^2}{v_F^2}\right)\right.\nonumber\\
&+&\left._2F_1\left(\frac{3}{2},\frac{1}{2};1;1-\frac{v_\Delta^2}{v_F^2}\right)\right],\,v_F\geq
v_\Delta,
\end{eqnarray}
and $_2F_1(a,b;c;x)$ is the hypergeometric function. If $v_\Delta
> v_F$, the two velocities should be exchanged. In the presence of
an $hc/2e$-vortex  the total energy of the system is thus
\begin{equation}
{\cal E}={\cal G}(R+l)-{\cal G}(R),
\end{equation}
where
\begin{equation}
{\cal G}(R)=-\int_{\Omega(R)}d^2{\bf
r'}\int_{\Sigma(\Lambda)}d^2{\bf k'} k'\rho(k',{\bf r'}).
\end{equation}
$\Omega(R)$ is the ellipse $v_F^2x'^{2}+v_\Delta^2y'^{2}\leq R^2$. To the leading
order in $1/(\Lambda R)$, one then finds ${\cal E}={\cal E}_0+{\cal E}_v$,
where ${\cal E}_0$ is the energy of the flux-free system given
by Eq.\ (\ref{vortex-free-anisotropic}), and the contribution to
the total energy arising solely from the presence of the AB vector potential when
$v_F \geq v_\Delta$ is
\begin{equation}\label{vortexE-anisotropic}
{\cal E}_v=\frac{\hbar lv_F}{12\pi R(R+l)}\,\,\,
{_2}F_1\left(\frac{1}{2},-\frac{1}{2};1;1-\frac{v_\Delta^2}{v_F^2}\right).
\end{equation}
In the isotropic case, $v_F=v_\Delta=1$, we obtain the result
(\ref{vortexE-isotropic}), since ${_2}F_1(1/2,-1/2;1;0)=1$. In the
opposite limit of a large velocity anisotropy, the energy cost is
determined by a larger of the two velocities, because the function
${_2}F_1(1/2,-1/2;1;x)$ is monotonic and bounded on the interval
$[0,1]$, $2/\pi={_2}F_1(1/2,-1/2;1;1)<{_2}F_1(1/2,-1/2;1;x)
<{_2}F_1(1/2,-1/2;1;0)=1$.

The result in Eq. (\ref{vortexE-anisotropic}) pertains to
the energy of filled quasiparticle
Fermi sea. However, we can straightforwardly import it into the fully
{\em self-consistent} computation of the {\em total} superconducting
condensation energy:
\begin{equation}\label{totalenergy}
E_{tot}(\Delta,hc/2e)=E_{qp}(\Delta,hc/2e)+\frac{|\Delta |^2}{g}~,
\end{equation}
where $E_{qp}(\Delta,hc/2e)$ is the energy of the Fermi see
with an $hc/2e$ vortex and gap parameter $\Delta$ and
$g$ is the effective coupling constant. We have
assumed that $\Delta$ is essentially uniform since any non-uniformity
enters only on the microscopic scale $\ll R$ or $l$.
By adding and subtracting
$E_{qp}(\Delta,0)$ we obtain:
\begin{equation}\label{totalenergydifference}
E_{tot}(\Delta,hc/2e)={\cal E}_v+E_{tot}(\Delta,0)~,
\end{equation}
where ${\cal E}_v$ is given by Eq. (\ref{vortexE-anisotropic}). By
minimizing $E_{tot}(\Delta,hc/2e)$ with respect to $\Delta$ we
obtain the total condensation energy in presence of $hc/2e$
vortex. This gives $\Delta=\Delta_0 + \delta\Delta$, where
$\Delta_0$ is the value that minimizes $E_{tot}(\Delta,0)$, and
$\delta\Delta\propto {\cal E}_v$. For macroscopic $R,l\gg
1/\Lambda$, ${\cal E}_v$ is arbitrarily smaller than
$E_{tot}(\Delta,0)$ and $\delta\Delta \ll \Delta_0$. This implies
that, to the leading order, the presence of an $hc/2e$ vortex
increases the condensation energy by precisely ${\cal E}_v$
(\ref{vortexE-anisotropic}), with $\Delta=\Delta_0$; the leading
correction is $\sim (\delta\Delta)^2/\Delta_0^2$.

For a finite s-wave gap $mR\ll 1$ the calculation is similar but considerably more cumbersome.
We find that the excess energy in Eq.\ (\ref{simple}) decreases with the gap $m$, and is essentially zero
already for $mR\approx 1$, which, crudely,
would correspond to an inner radius of a micrometer in an aluminum ring.
This is in accord with our interpretation of the effect as being due
to the nodal quasiparticles, and with
the high accuracy of the observed $hc/2e$-periodicity in standard low-$T_c$ superconductors.

We can estimate the above energy cost of the $hc/2e$-vortex from
the values of the Fermi velocity and the velocity anisotropy in
YBa$_2$Cu$_3$O$_{7-\delta}$ (YBCO) and
Bi$_2$Sr$_2$CaCu$_2$O$_{8+y}$ (BSCCO) obtained by ARPES
\cite{Zhou} and the thermal conductivity measurements
\cite{Chiao}. ARPES yields a value of the Fermi velocity
$v_F\sim3\cdot10^5$m/s, which appears to be universal in cuprates.
The velocity anisotropy is $v_F/v_\Delta\sim 14$ in YBCO,  while
in BSCCO $v_F/v_\Delta\sim 19$, yielding the total energy penalty
of having an $hc/2e$-vortex, ${\cal E}_v^{\rm tot}=N{\cal E}_v\sim
0.2$K, for an annulus with the inner radius $R=1\mu$m, and $N=4$
as the number of nodes in a d-wave superconductor.

The finite energy cost of an $hc/2e$-vortex will affect the
quantization of the magnetic flux when it becomes comparable to
the second relevant energy scale in the problem, namely the
superfluid density, $\rho(T)$. By lifting the parabolas centered
at the $hc/2e$ flux in the text-book energy vs. magnetic flux plot
\cite{doll} by ${\cal E}_v$ it is easy to see that the width of
the $hc/e$ relative to the one of the $hc/2e$ plateau becomes
longer by an amount proportional to $\delta={\cal E}_v / \rho(T)$.
This is typically a small number: in optimally doped YBCO, for
example, $\delta\approx 10^{-4}$. Recently, however, single
crystals \cite{broun} and thin films \cite{lemberger} of severely
underdoped YBCO have been studied with the unprecedented low
$\rho(0)\sim 1K$, when expressed in energy units \cite{case}. This
extremely underdoped regime where the phase stiffness can be
rendered arbitrary small with underdoping offers the best chance
for an observation of the asymmetry between even- and odd-flux
vortices predicted in this paper.

 The asymmetry between $hc/e$- and $hc/2e$- vortices in unconventional superconductor has also been recently
 found in the numerical solution of the Bogoliubov - de Gennes equations for a mesoscopic superconducting loop
 \cite{loder} (see also \cite{czajka}). The results and conclusions of this work seem broadly consistent with ours.

 The authors are grateful to D. Bonn, M. Franz, J. Kirtley, K. Moller, and O. Vafek,  for useful discussions,
 and the Aspen Center for Physics where this work was initiated.
 V. J. and I. F. H. acknowledge the support from the NSERC of Canada, and Z. T. from  NSF grant DMR-0531159.

%%%%%%%%%%%%%%%%%%%%%%%%%%%%%%%%%%%%%%%%%%%%%%%%%%%%%%%%%%%%%%%%%%

\end{document}